\documentclass[traditabstract]{aa}
\usepackage{bm}
\usepackage{graphicx}
\usepackage{amsfonts}
\usepackage{textcomp}
\usepackage{txfonts}

\begin{document}%opening
\title{$J$-state interference signatures in the Second Solar Spectrum}
\subtitle{Modeling the Cr\,{\sc i} triplet at 5204-5208\,\AA \\}
\author{H. N. Smitha
\inst{1}
\and K. N. Nagendra
\inst {1}
\and J. O. Stenflo
\inst{2,3}
\and M. Bianda
\inst{3}
\and M. Sampoorna
\inst{1}
\and R. Ramelli
\inst{3}
\and L. S. Anusha
\inst{1}}
\institute{Indian Institute of Astrophysics, Koramangala, Bangalore, India
\and Institute of Astronomy, ETH Zurich, CH-8093 \  Zurich, Switzerland 
\and Istituto Ricerche Solari Locarno, Via Patocchi, 6605 Locarno-Monti, Switzerland
\date{}}
%\shortauthors{Smitha et al.}

\abstract
{The scattering polarization in the solar spectrum 
is traditionally modeled with each spectral line treated separately,
but this is generally inadequate for multiplets where $J$-state interference plays a significant role.
Through simultaneous observations of all the 3 lines of a Cr\,{\sc i} triplet, combined with realistic 
radiative transfer modeling of the data, we show that it is necessary to include
$J$-state interference consistently when modeling lines with partially interacting 
fine structure components. Polarized line formation theory that includes $J$-state 
interference effects together with partial frequency redistribution for a two-term atom 
is used to model the observations. Collisional frequency
redistribution is also accounted for. We show that the resonance polarization in 
the Cr\,{\sc i} triplet is strongly affected by the partial frequency redistribution
 effects in the {line core}  and near wing peaks. The Cr\,{\sc i} triplet is quite 
sensitive to the temperature structure of the photospheric layers.
{Our complete frequency redistribution calculations in semi-empirical models 
of the solar atmosphere cannot reproduce the observed near wing polarization or the
cross-over of the Stokes $Q/I$ line polarization about the continuum polarization level
that is due to the $J$-state interference. When however partial frequency 
redistribution is included, a good fit to these features can be achieved.
Further, to obtain a good fit to the far wings, a small temperature 
enhancement of the FALF model in the photospheric layers is necessary.} 
}


\maketitle

%=============================================================================
\section{Introduction}
The Second Solar Spectrum is produced by coherent scattering processes
in the solar atmosphere (Stenflo 1994). 
Modeling this spectrum requires the solution of the polarized radiative transfer
(RT) equation. It is well known that quantum interference
between the fine structure ($J$) levels is responsible for the
formation of line pairs such as Na\,{\sc i} D$_1$ and D$_2$,
Ca\,{\sc ii} H and K, etc. The linear polarization in the Ca\,{\sc ii} H and K lines was
observed and modeled by Stenflo (1980), who developed a theoretical
framework which for the first time demonstrated the
profound role that quantum 
interference between different fine structure components within a
multiplet can have on the observed polarization profiles. 
An extended version of this theoretical framework, applicable to any multiplet,
was presented in Stenflo (1997). It was used for RT modeling of 
$J$-state interference in the Na\,{\sc i} D$_1$ and D$_2$ lines by Fluri et al.~(2003).
The $J$-state interference theory used in the mentioned 
papers assumed frequency coherent scattering. Recently 
Smitha et al.~(2011a, hereafter P1) have extended the theory of Stenflo (1997)
to include partial frequency redistribution (PRD) with $J$-state interference.
It is restricted to the case of a two-term atom and uses the
assumption that the lower term is unpolarized. Landi Degl'Innocenti et
al.~(1997) have proposed
an alternative approach to the same problem based on the concept
of metalevels, which is intended for dealing with PRD 
in multi-term atoms. A multi-term formulation of 
$J$-state interference that accounts for the polarization
 in all the terms, but which is restricted to the case of 
complete frequency redistribution (CRD), has been given in 
Landi Degl'Innocenti \& Landolfi  (2004, hereafter LL04).
 In the mentioned theoretical formulations 
collisional frequency redistribution was not included. 

In Smitha et al.~(2011b, hereafter P2) the redistribution matrix 
for the $J$-state interference derived in P1 was 
incorporated into the RT equation, which was solved for simple
isothermal model atmospheres. Several theoretical aspects 
of radiative transfer in a hypothetical doublet line system are
studied in that paper. 
The purpose of the present paper is to perform {one-dimensional}
 RT modeling of the polarimetric observations of a multiplet 
where $J$-state interference is relevant. For this we have 
selected the Cr\,{\sc i} triplet at 5204.50\ \AA\ (line-1: $J_b=1 \to J_a=2$), 
5206.04\ \AA\ (line-2: $J_b=2 \to J_a=2$), and 5208.42\ \AA\ (line-3: $J_b=3 \to J_a=2$). 
Hyperfine splitting can be neglected because the most abundant (90\%) 
isotope of Cr\,{\sc i} has zero nuclear spin. 

Kleint et al.~(2010a,b) have 
used the Cr\,{\sc i} triplet for a synoptic program to explore solar 
cycle variations of the microturbulent
field strength. {Recently, Belluzzi \& Trujillo Bueno (2011) 
applied the density matrix theory described in LL04 
(which is based on the CRD
approximation) in order to perform a basic investigation 
on the impact of $J$-state interference in several important multiplets in the 
solar spectrum including also the Cr\,{\sc i} triplet.} 
In this work they have neglected RT and PRD effects. 
However, they have included the effects of lower term 
polarization and the dichroism. They identify 
and explain qualitatively the observational signatures 
produced by $J$-state interference in the Cr {\sc i} 
triplet (i.e., the cross-over of $Q/I$ about the continuum 
polarization level occurring between the lines, and the $Q/I$
feature around the line-1 core).

 In Section~{\ref{rt}} we briefly present the basic equations
required for realistic RT modeling of lines in the two-term 
atom picture. In Section~{\ref{obs}} we present the 
polarimetric observations of the Cr\,{\sc i} triplet.
 Section~{\ref{model}} is devoted to a description of 
realistic modeling of the observations. In Section~{\ref{results}}
 we present the main results. Concluding remarks are given in 
Section~{\ref{conclusions}}. 

%=========================================================================
\section{Polarized line transfer equation for a two-term atom}
\label{rt}
\begin{figure}
\centering
\includegraphics[width=8.5cm, height=5.0cm]{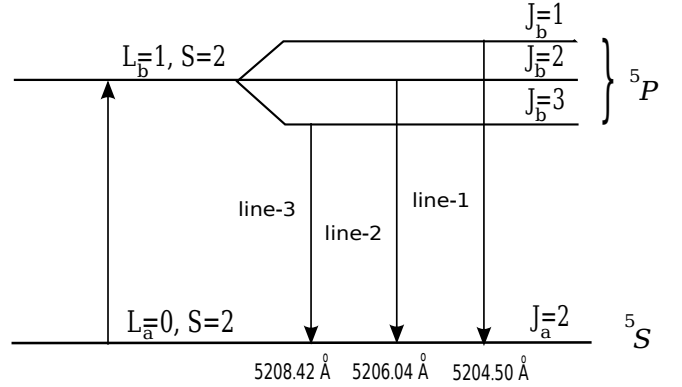}
\caption{The term diagram showing transitions in the Cr I triplet. 
The diagram is not drawn to scale.}
\label{level-diag}
\end{figure}

In a non-magnetic medium, the polarization of the radiation field
is represented by the Stokes vector $(I,Q)^{\rm T}$, where
positive $Q$ is defined to represent linear polarization that is
oriented parallel to the solar limb. 
In a medium that is axisymmetric around the vertical direction, it is
advantageous to use a formulation in terms of the 
reduced Stokes vector ${\bm {\mathcal{I}}}$ instead of the
traditional Stokes vector $(I,Q)^{\rm T}$. 
The transformations between the two can be found in Appendix~B of Frisch (2007).
The relevant line transfer equation 
for the 2-component reduced Stokes vector is
\begin{eqnarray}
\mu \frac{{\partial {\bm{\mathcal I}}(\lambda, \mu, z)}
}{\partial z}= 
-k_{\rm tot}(\lambda, z)
\left[{\bm{\mathcal I}}(\lambda, \mu, z) - 
{\bm{\mathcal S}}(\lambda, z)\right],
\label{transfer}
\end{eqnarray}
in standard notation (see Anusha et al.~2011).
$z$ is the geometric height in the atmosphere. See P2
for details of Eq.~(\ref{transfer}) and related quantities. 
The total opacity 
$k_{\rm tot}(\lambda, z)= \eta_{M}(\lambda,z) + 
\sigma_c(\lambda, z) + k_c(\lambda, z)$, where  $\sigma_c$ and 
$k_c$ are the continuum scattering and continuum absorption
coefficients, respectively. The line absorption coefficient for the entire 
multiplet is
\begin{eqnarray}
\eta_{M}(\lambda,z)\! &=&\!k_{M}(z)\phi_{M}(\lambda,z)
=\!\!\sum_{J_a J_b} k_{l(J_bJ_a)} \phi(\lambda_{J_bJ_a},z),
\label{comb-prof-func}
\end{eqnarray}
where $k_{l(J_bJ_a)}$ is the wavelength averaged absorption coefficient 
for the $J_a \to J_b$ transition with the corresponding profile
function denoted by $\phi(\lambda_{J_bJ_a},z)$. 
$J_b$ and $J_a$ are the
total angular momentum quantum numbers of the fine-structure levels 
for the upper and lower terms (see Fig.~{\ref{level-diag}}). 
$k_{M}(z)$ is the wavelength averaged absorption 
coefficient for the entire multiplet.
For our case of a two-term atom, we need to define a 
combined profile function that determines the shape of the absorption 
coefficient across the whole multiplet. It can be shown that for 
the Cr\,{\sc i} triplet line system $\phi_M(\lambda,z)$ is given by
(see Eqs.~(7) and (8) in 
Section~2 of P2)

\begin{eqnarray}
\phi_{M}(\lambda,z)=\frac{3 \phi(\lambda_{1\,2},z)+5\phi(\lambda_{2\,2},z)
+7\phi(\lambda_{3\,2},z)}{15}.
\end{eqnarray}
The reduced source vector is defined as
\begin{eqnarray}
\bm{\mathcal S}(\lambda, z)&=&\frac{k_M(z)\phi_{M}(\lambda, z)
\bm{\mathcal S}_{l}(\lambda, z)+\sigma_c(\lambda, z) \bm{\mathcal S}_{c}(\lambda, z)}
{k_{\rm tot}(\lambda, z)} \nonumber \\ &&
+\frac{ k_c(\lambda, z) \bm{\mathcal S}_{\rm th}(\lambda, z)}
{k_{\rm tot}(\lambda, z)},
\label{s-total}
\end{eqnarray}
for a two-term atom with an unpolarized lower term. Here
$\bm{\mathcal S}_{\rm th}=(B_{\lambda},0)^T$, where $B_{\lambda}$ is
the Planck function.
The continuum scattering source vector is
\begin{equation}
{\bm{\mathcal  S}}_{c}(\lambda, z)= \frac{1}{2}\int_{-1}^{+1} \hat\Psi(\mu')
{\bm{\mathcal I}}(\lambda, \mu', z)\,d\mu'.
\label{irreducible-sc}
\end{equation}
Since continuum polarization can be seen as representing
scattering in the distant wings of spectral lines (in particular
from the Lyman series lines, cf. Stenflo 2005), we are justified
to use the assumption of frequency coherent scattering for the
continuum.
The matrix $\hat\Psi$ is the Rayleigh scattering phase matrix in the 
reduced basis (see Frisch 2007). The line source vector 
\begin{eqnarray}
{\bm{\mathcal  S}}_{l}(\lambda, z)&=&
\epsilon\bm{\mathcal S}_{\rm th}(\lambda, z)+
\int_{-\infty}^{+\infty}\frac{1}{2}\nonumber \\ && \ \ \ \ \ \ \ 
\times\int_{-1}^{+1} 
\frac{\widetilde{{\bm {\mathcal R}}}(\lambda ,\lambda^{\prime}, z)}
{\phi_M(\lambda, z)}  \hat\Psi(\mu') {\bm{\mathcal I}}
(\lambda', \mu', z)\,d\mu'\,d\lambda'.
\label{irreducible-sl}
\end{eqnarray}
%%%%%%%%%%%%%%%%%%%%%%%%%%%%%%%%%%%%%%%%%%%%%%%fig1
%%%%%%%%%%%%%%%%%%%%%%%%%%%%%%%%%%%%%%%%%%%%%%%fig1
 %$\epsilon$ is defined by
The thermalization parameter $\epsilon={\Gamma_I}/(\Gamma_R+\Gamma_I)$ 
where $\Gamma_R$ and $\Gamma_I$ are
the radiative and inelastic collisional de-excitation rates, respectively. 
$\widetilde{{\bm {\mathcal R}}}(\lambda,\lambda^{\prime}, z)$ 
is a $(2\times 2)$ diagonal matrix with elements
 $\widetilde{{\bm {\mathcal R}}}=$ diag $({\mathcal R}^{0},{\mathcal R}^{2})$, 
where ${\mathcal R}^{K}$ are the redistribution 
functions which include the effects of $J$-state interference between different line 
components in a multiplet.
{${\mathcal R}^{K}$ represents a linear combination of redistribution 
functions of type-II and type-III.
In the reduced Stokes vector basis, the angular phase matrix and the frequency
redistribution functions are decoupled. The phase matrix part is built into the
transfer equation through the $\hat\Psi$ matrix.
The theory of redistribution matrices for the $J$-state interference
in two-term atoms for the collisionless case is presented in P1.
This frequency redistribution part that includes $J$-state interference and
the collisional redistribution is given by

\begin{eqnarray}
&&{\bf {\mathcal R}}^{K}(x,x^\prime) = \frac{3(2L_b+1)}{2S+1}
\sum_{J_aJ_fJ_bJ_{b^\prime}} (-1)^{J_f-J_a} \nonumber \\ && \times
  \langle \tilde r_{J_b} \tilde r^\ast_{J_{b^\prime}} 
\rangle_{J_aJ_f} (2J_a+1)
(2J_f+1)(2J_b+1)(2J_{b^\prime}+1) \nonumber \\&& \times 
\left\lbrace 
\begin{array}{ccc}
L_a & L_b & 1\\
J_b & J_f & S \\
\end{array}
\right\rbrace
\left\lbrace 
\begin{array}{ccc}
L_a & L_b & 1\\
J_b & J_a & S \\
\end{array}
\right\rbrace 
\left\lbrace 
\begin{array}{ccc}
L_a & L_b & 1\\
J_{b^\prime} & J_f & S \\
\end{array}
\right\rbrace \nonumber \\ && \times
\left\lbrace 
\begin{array}{ccc}
L_a & L_b & 1\\
J_{b^\prime} & J_a & S \\
\end{array}
\right\rbrace 
\left\{
\begin{array}{ccc}
1 & 1 & K \\
J_{b^\prime} & J_b & J_a \\
\end{array}
\right\}
\left\{
\begin{array}{ccc}
1 & 1 & K \\
J_{b^\prime} & J_b & J_f \\
\end{array}
\right\}.
\label{r_mat}
\end{eqnarray}
The ensemble averaged coherency matrix elements (see e.g., Bommier \& Stenflo 1999)
in the above equation are given by
\begin{eqnarray}
&&\langle \tilde r_{J_b} \tilde r^\ast_{J_{b^\prime}} 
\rangle_{J_aJ_f} = A  \cos\beta_{J_{b^\prime}J_b} \nonumber \\ && \times [\cos\beta_{J_{b^\prime}J_b}
(h^{\rm II}_{J_b,J_{b^\prime}})_{J_aJ_f} -\sin\beta_{J_{b^\prime}J_b}
(f^{\rm II}_{J_b,J_{b^\prime}})_{J_aJ_f}] %_{\rm \Theta avg}
\nonumber \\ && + B^{(K)} \cos\beta_{J_{b^\prime}J_b}\cos\alpha^{(K)}_{J_{b^\prime}J_b}
 \nonumber \\ &&\!\!\!\! \times 
\Bigg\{\cos\big(\beta_{J_{b^\prime}J_b}+ \alpha^{(K)}_{J_{b^\prime}J_b}\big) 
\left[\Re\bigg({h^{\rm III}_{J_bJ_a,J_{b^\prime}J_f}}\bigg)
-\Im{\bigg(f^{\rm III}_{J_bJ_a,J_{b^\prime}J_f}\bigg)}\right]\nonumber \\ && \!\!\!\!-
\sin\big(\beta_{J_{b^\prime}J_b}+ \alpha^{(K)}_{J_{b^\prime}J_b}\big)
\left[\Im\bigg({h^{\rm III}_{J_bJ_a,J_{b^\prime}J_f}}\bigg)
+\Re{\bigg(f^{\rm III}_{J_bJ_a,J_{b^\prime}J_f}\index{}\bigg)}\right]
\Bigg\}.%_{\rm \Theta avg}.\nonumber \\
\label{coherency_mat_lab}
\end{eqnarray}
\noindent
The $(h^{\rm II}_{J_b,J_{b^\prime}})_{J_aJ_f}$ and 
$(f^{\rm II}_{J_b,J_{b^\prime}})_{J_aJ_f}$ are the auxiliary functions defined in
Eqs.~(14) and (15) of P1 but are used here for the non-magnetic case
and with angle-averaged redistribution functions of type-II.
The auxiliary functions of type-II derived in P1
represents generalizations of the corresponding
quantities appearing in Sampoorna (2011, see Eqs.~(22) and (23))
using a semi-classical approach. The important difference between the two in the
presence of the magnetic field is that in case of $J$-state interference
these auxiliary functions depend on both $J$ and $m$ 
quantum numbers, unlike in case of
$m$-state interference where they depend only on $m$.
In the particular case of non-magnetic $J$-state interference
theory, these quantities depend only on $J$ quantum numbers, whereas
the corresponding quantities in the $m$-state interference 
theory simply reduce to the well known type-II redistribution
functions of Hummer (1962). Therefore for the notational 
brevity even in the $J$-state interference 
case we refer to these auxiliary functions as $R_{\rm II}$ hereafter
(in the standard notation of Hummer 1962).

The auxiliary quantities for type-III redistribution (see the last two lines of 
Eq.~(\ref{coherency_mat_lab}))
in the non-magnetic case are also derived in a 
way similar to the two-level atom case given in Eqs.~(27)-(36) of Sampoorna (2011).
But in doing so the following assumptions are made
\begin{enumerate}
 \item Infinitely sharp lower term. 
\item Unpolarized lower term.
\item Weak radiation field limit (i.e., stimulated emission is neglected 
in comparison with spontaneous emission).
\item Hyperfine structure is neglected.
\item The (inelastic) collisions that transfer polarization between 
fine structure levels of a given term (upper or lower) are neglected.
\item The depolarizing elastic collisions that couple $m$-states 
belonging to a given fine structure level $J_b$ are considered
and taken to be same for a given term.
\item We restrict our attention to the linear Zeeman regime
(Zeeman splitting much smaller than the fine structure splitting).
This is not applicable in the Paschen-Back regime.
\end{enumerate}

The assumption of an unpolarized lower term is made for the sake of 
mathematical simplicity.
The inelastic collisions that transfer polarization between the fine
structure levels are neglected.
 This is justified because the colliding particles are isotropically 
distributed around the radiating atom. This situation is similar 
to the case of an atom immersed in an isotropic radiation field 
producing no atomic polarization (scattering polarization). 
Neglecting such inelastic collisions is particularly 
valid in the linear Zeeman regime in which
we are interested. However these inelastic
collisions do cause population
transfer between the fine structure levels, 
and are properly accounted for in the calculations of line opacities 
(see Section~{\ref{model}}).

The angle $\beta_{J_{b^\prime}J_b}$
is defined in Eq.~(10) of P1.
The angle $\alpha^{(K)}_{J_{b^\prime}J_b}$ is defined as
\begin{equation}
\tan \alpha^{(K)}_{J_{b^\prime}J_b}= \frac{\omega_{J_{b^\prime}J_b}} 
{\Gamma_{R}+\Gamma_{I}+D^{(K)}}.
\label{alpha-k}
\end{equation}
Here $\hbar \omega_{J_{b^\prime}J_b}$ is the energy difference between
the $J_{b^\prime}$ and $J_b$ states in the absence of a magnetic
field. $D^{(K)}$ is the $2K$ multipole depolarizing elastic collisional 
destruction rate.
In general $D^{(K)}$ depend on the
$J$ quantum numbers of the fine structure states. 
As a simplifying assumption, we take these rates to
be the same for all the fine structure states of the
upper term, and use the classical value
$D^{(K)} = \Gamma_E/2$ given by Stenflo (1994) where $\Gamma_E$
is the elastic collision rate which is responsible for the 
broadening of the atomic states.
$A$ and $ B^{(K)}$ are the branching ratios which are given by 
\begin{eqnarray}
 A=\frac{\Gamma_{R}}{\Gamma_{R}+\Gamma_{I}+\Gamma_{E}};
\quad B^{(K)}=\frac{\Gamma_{R}}{\Gamma_{R}+\Gamma_{I}+D^{(K)}}\frac{\Gamma_{E}-D^{(
K)}}{\Gamma_{R}+\Gamma_{I}+\Gamma_{E}}.
\label{a-bk}
\end{eqnarray}
These branching ratios are the ones derived for a two-level atom model
by Bommier (1997). In view of the simplifying assumptions stated above,
we continue to use the same branching ratios for the two-term atom case also. 

The computation of angle-averaged type-III redistribution functions 
(that appear in Eq.~(\ref{coherency_mat_lab}))
is very expensive, especially for the case
of a realistic model atmosphere. Therefore we prefer to use
the approximation of CRD in place of type-III redistribution functions (see Mihalas 1978).
We have verified by direct numerical computations that this replacement is valid
and gives results which are almost identical to the explicit
use of type-III redistribution functions. The necessary settings of the branching 
ratios in order to go to the limit of pure CRD are $A=0$ and $B^{(K)}=(1-\epsilon)$ 
(see also Anusha et al. 2011). }
%===========================================================================
\section{Observational details}
\label{obs}
\begin{figure}
\centering
\includegraphics[width=9.0cm, height=8.0cm]{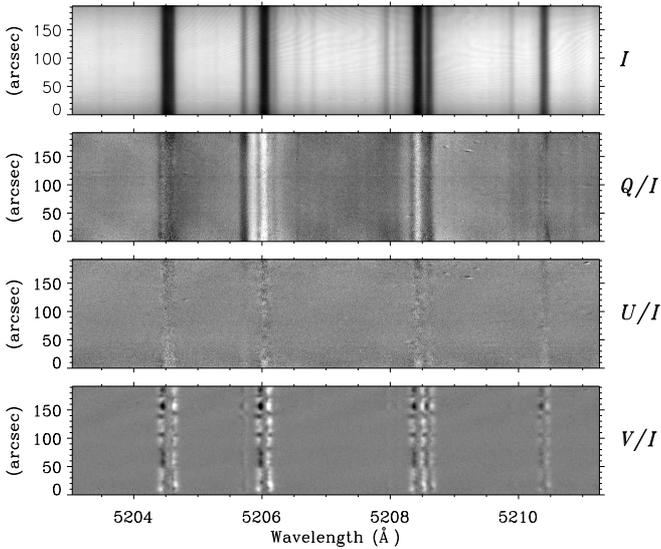}
\caption{CCD image
showing the $(I, Q/I, U/I, V/I)$ of the Cr\,{\sc i} triplet.
The recording was made on  September 6, 2011, near the heliographic
north pole at a limb distance defined by $\mu =0.15$.  
The grey scale cuts span a range (from black to white) of
0.1\,\%\ in $Q/I$ and $U/I$, while for $V/I$ the cuts are $-0.2$\,\%\ (black) and
$+0.2$\,\%\ (white).}
\label{observations}
\end{figure}

{The $Q/I$ spectra of Cr\,{\sc i} triplet were observed by Gandorfer (2000).
In this paper, we present new observations of this triplet} 
obtained with the ZIMPOL-3 polarimeter 
(Ramelli et al.~2010) at IRSOL in Switzerland.
Fig.~{\ref {observations}} shows the observations recorded on September 6,
2011, at the heliographic north pole with the slit placed 
parallel to the limb at $\mu=0.15$. The polarization modulation was
done with a piezo-elastic modulator (PEM).
The spectrograph slit was $60\mu$m wide corresponding to a spatial 
extent of $0.5^{\prime\prime}$ on 
the solar image. The CCD covered $190^{\prime\prime}$ along
the slit.
%The spectrographic slit was $60\mu$m wide corresponding to a
%spatial extent of $0.5^{\prime\prime}$ across,
% and $190^{\prime\prime}$ along on the disk.
The effective pixel resolution in the spatial direction is 4 actual
pixels wide, due to the grouping of each four pixel rows covered by a
cylindrical microlense, to allow simultaneous
recording of all four Stokes parameters in the ZIMPOL demodulation
scheme. The resulting CCD images have 140 
such effective pixel resolution elements 
in the spatial direction, each element corresponding to $1.38^{\prime\prime}$,
and 1240 pixels in the wavelength direction, with one pixel 
corresponding to 7.84 m\AA. 
In Fig.~{\ref{observations}} only the central part of the spectral
window, corresponding to 1050 pixels
spanning 8.23\ \AA, is displayed.
With the PEM it was possible to measure simultaneously one linear 
and the circular polarization component. Measurements 
of the linear polarization component $Q/I$ were alternated 
with measurements of the $U/I$ component through mechanical rotation
of the analyzer by $45^{\circ}$. In total we accumulated for both 
the components 2000 exposures of 1 second each.
%%%%%%%%%%%%%%%%%%%%%%%%%%%%%%%%%%%%%%%%%%%%%%%fig1
%%%%%%%%%%%%%%%%%%%%%%%%%%%%%%%%%%%%%%%%%%%%%%%fig1
\section{Modeling of Cr\,{\sc i} triplet}
\label{model}
To model the Cr\,{\sc i} triplet the polarized spectrum
is calculated by a two-stage process 
described in Holzreuter et al.~(2005, see also Anusha et al.~2010, 2011). 
In the first-stage a multi-level PRD-capable MALI (Multi-level Approximate 
Lambda Iteration) code of Uitenbroek (2001, referred to as the RH-code) solves the 
statistical equilibrium equation and the unpolarized RT
equation self-consistently and iteratively.
The RH-code is used to compute the unpolarized intensities, opacities and the 
collision rates. The angle-averaged redistribution functions of 
Hummer (1962) are used in the RH-code to represent
PRD in line scattering. In the second stage  the 
opacities and the collision rates are kept fixed, and the polarized 
intensity vector $\bm{\mathcal I}$ is computed perturbatively by 
solving the polarized RT equation with the redistribution 
matrices defined in Section~{\ref{rt}},
which are derived for a two-term atom with an
unpolarized lower term. 

%==============================================================================
\subsection{Model atom and model atmosphere}
\label{model-atm}
The Cr\,{\sc i} atom model is constructed for 14 
levels {(13 levels of Cr\,{\sc i} and the ground state of Cr\,{\sc ii})},
 11 line transitions, and 13 continuum transitions. 
The line components of the $^5\!S-^5\!\!P$ triplet of Cr\,{\sc i} 
are considered under PRD. The values of the various
physical quantities required to build the atom model are
taken from the NIST atomic data base\footnote[1]
{www.nist.gov/pml/data/asd.cfm} and the Kurucz 
data base\footnote[2]{kurucz.harvard.edu/linelists.html}. 
The data for the blend lines are also obtained from the Kurucz data base.
The photo-ionization cross sections are taken from 
Bergemann \& Cescutti (2010). The explicit dependence
of these cross sections on wavelength is computed 
under the hydrogenic approximation. 

\begin{figure}
\centering
\includegraphics[width=9.0cm, height=12.0cm]{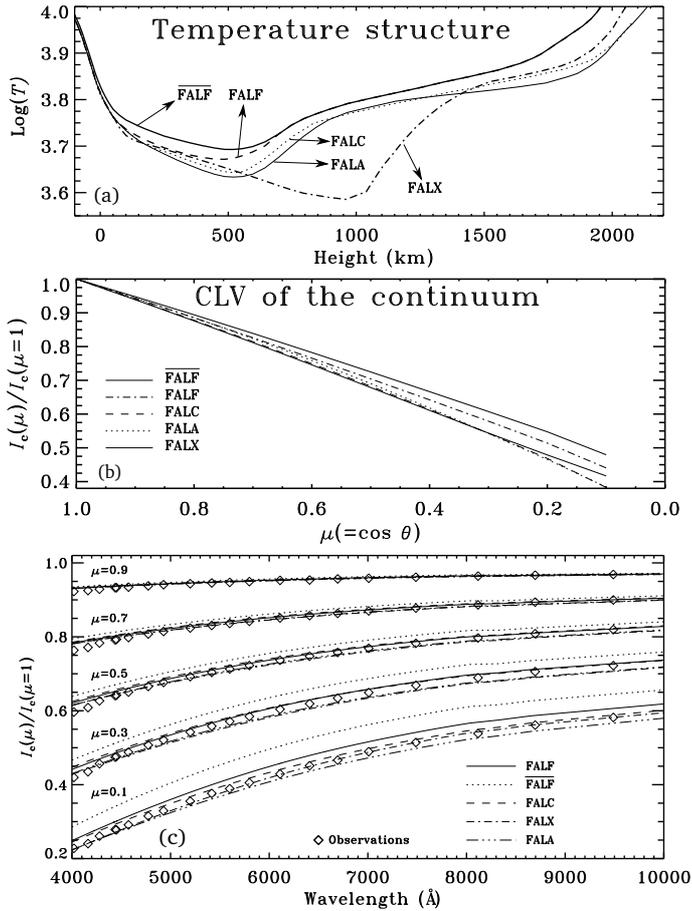}
\caption{ Panel (a) shows the temperature structure of several 
standard model atmospheres. $\overline{\rm FALF}$ represents a 
model with an enhanced temperature structure of the original
FALF model. In panel (b) we show the center to limb variation 
of the `limb darkening function', $I_c(\mu)/I_c(\mu=1)$, where 
$I_c(\mu)$ is the continuum intensity near the Cr\,{\sc i} triplet.
{Panel (c) shows the CLV of the continuum intensity for all
wavelengths covering the violet to the IR regions of the spectrum. 
The observed data are taken from Neckel \& Labs (1994).}}
\label{clv-temp}
\end{figure}

Fig.~{\ref{clv-temp}a} shows the temperature structure in some
of the standard model atmospheres of the Sun - namely
FALA, FALC, FALF (Fontenla et al. 1993), and FALX (Avrett 1995), 
which we have used in our attempts to fit the observed $(I, Q/I)$ spectra. Models
A, C and F of Fontenla et al. (1993) represent respectively
the supergranular cell center, the average quite Sun, and the
 bright network region in the solar atmosphere. FALX represents 
the coolest model with a chromospheric temperature minimum located 
around 1000 km above the photosphere. Our attempts to fit the 
observed $(I, Q/I)$ spectra using these standard models will be discussed in Sect.~5.2.

We show that a reasonable fit could be obtained only with a
temperature structure modified with a small enhancement 
in the original temperature structure of the FALF model,
in the height range of 100 km below the photosphere, extending up to 700 km above 
the photosphere (denoted by $\overline{\rm FALF}$). Such an 
enhancement does not affect the center to limb variation (CLV) 
of the continuum intensity as shown in Fig.~{\ref{clv-temp}b}. 
While such a modification of the temperature structure produces 
insignificant changes in the intensity spectra, $Q/I$ turns out 
to be quite sensitive to these changes in the temperature gradient.

{In order to explore the effect of temperature enhancement
in a given model atmosphere used for computing line and 
continuous spectra, we have performed a simple test (similar to the Fig.~3 of
Asplund et al. 1999, see our Fig.~{\ref{clv-temp}c}).
It is expected that 1D model atmospheres (like FALF in our case; or
all the FAL class of models in general) fit the observed CLV data 
of continuum intensity to a good accuracy. 
To verify this, we have plotted the limb darkening function for the whole range 
of wavelengths, for different $\mu$ values. The theoretical
limb darkening function fits the corresponding observed data better for $\mu \to 1$.
The fit is approximate in the limb positions (say $\mu=0.1$).
In Fig.~{\ref{clv-temp}c} we also show the theoretical
curves computed for the $\overline{\rm FALF}$ model (dotted lines).
As can be seen, the limb darkening function of $\overline{\rm FALF}$
does not greatly differ from that of the original FALF model atmosphere
(a maximum relative difference of 15\% in the extreme limb).
Therefore it justifies a slight modification of temperature structure 
in a given model atmosphere to achieve a theoretical fit to the $Q/I$
observations.}

%-----------------------------------------------------

%-==============================================================================
\section{Results and discussion}
\label{results}
%===============================================================================
\subsection{Comparison between PRD and CRD}
\label{crd-prd-theo}
\begin{figure}
\centering
\includegraphics[width=9.0cm, height=7.0cm]{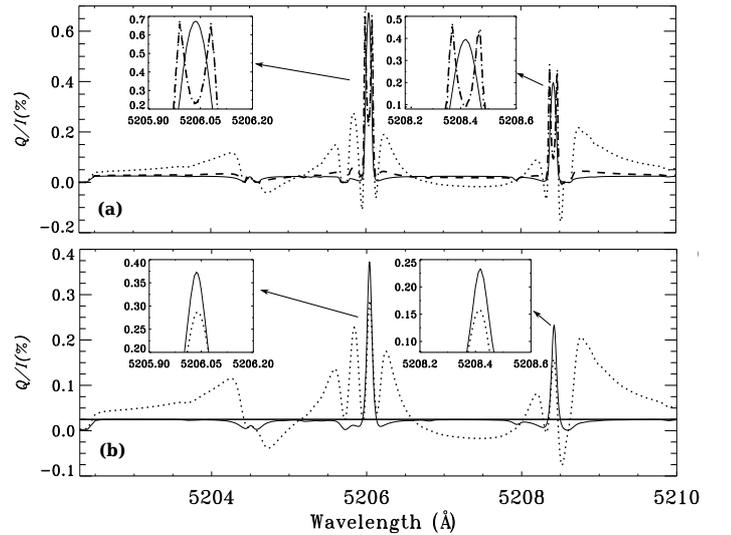}
\caption{\footnotesize {($Q/I$) of the Cr\,{\sc i} 
triplet computed with CRD (solid line),
$R_{\rm II}$ (dotted line), and a combination of CRD and $R_{\rm II}$
(dashed line) {for $\mu=0.15$. 
The positive $Q$ represents linear polarization
parallel to the solar limb.} %The model atmosphere used is a modified version of FALF. 
The thin solid line in $Q/I$ at the 0.025\,\%\ level represents the 
continuum polarization.   
The microturbulent magnetic field $B_{\rm turb}=0$. No spectral smearing
is applied to the profiles in the panel~(a).
In the line core the $Q/I$ profiles computed with 
pure $R_{\rm II}$ and with a combination of $R_{\rm II}$
and CRD nearly coincide (see the insets in the panel~(a)). 
The panel~(b) shows a comparison between the $Q/I$
profiles computed with CRD and with $R_{\rm II}$ when we also 
apply a spectral smearing of 80\,m\AA. 
The line types have the same meanings as in panel~(a).}
}
\label{crd-prd}
\end{figure}

Fig.~{\ref{crd-prd}} shows the comparison between the $Q/I$
profiles computed using only CRD (to represent frequency non-coherent
redistribution: solid line),
only $R_{\rm II}$ (dotted line), and a combination of 
$R_{\rm II}$ and CRD (dashed line) {(see Section~\ref{rt} for the definition of CRD)}.
As seen from the figure,
the CRD profiles do not produce the wing peaks 
on either side of the line center, which are clearly seen in the
PRD profiles. Also, the $J$-state interference signatures are more prominent
in PRD than in CRD. A good fit to the observed $Q/I$ can only be achieved 
through the use of PRD (see Section~{\ref{obs-theo}}).
We have verified that it is
impossible to fit the observed near wing peaks with CRD alone.
The pure $R_{\rm II}$ mechanism represents frequency coherent scattering in the
line wings, the use of which alone also fails to achieve a good fit 
(since it produces too large values of $Q/I$ throughout the wings). We find
that a proper combination of $R_{\rm II}$ and CRD is essential
to obtain a good fit to the observations.
This can be seen more clearly in Fig.~{\ref{fit}}, where we present 
a comparison with the observed $Q/I$ profile.
It is well known that only 
such a combination can correctly 
take into account the collisional frequency redistribution. 

Therefore the $(R_{\rm II}, $CRD$)$ combination is adopted for the computations 
in the present paper. 

The effect of elastic collisions is to cause significant depolarization
in the line wings. 
This can be seen from the dashed line in Fig.~{\ref{crd-prd}a},
which shows that due to elastic collisions the line wing amplitudes
of $Q/I$ are greatly suppressed with respect to the corresponding
pure $R_{\rm II}$ case (dotted line).
The fact that the Cr\,{\sc i} line components are formed
in the upper photosphere and lower chromosphere makes it necessary 
to include elastic collisions in our theoretical modeling.
The issue of elastic collisions is discussed in some detail in {Section~{\ref{rt}}}.

%---------------------------------------------------------------------------------
%%%%%%%%%%%%%%%%%%%%%%%%%%%%%%%%%%%%%%%%%%%%%%%fig3

Fig.~{\ref{crd-prd}b} shows the effect 
of spectral smearing that needs to be applied to the theoretical
profiles in order to compare them with the observed profiles, which
are broadened by the particular spectral resolution that was used in
the observations. In the absence of smearing the $Q/I$ 
in the line core computed with $R_{\rm II}$ and CRD differ significantly 
(see Fig.~{\ref{crd-prd}a}).
These differences decrease drastically after application of spectral smearing.
Although in isothermal atmospheric models $Q/I$ computed with pure
$R_{\rm II}$ and with CRD are very similar in the line center region, the same cannot be 
expected in computations with realistic atmospheres.
Indeed $Q/I$ computed with PRD shows a double peak structure in the line
core region with a dip at line center (see Holzreuter et al. 2005 for details).
The smearing wipes out the double-peaked core 
structure that we see in Fig.~{\ref{crd-prd}a}.

%====================================================================================
\subsection{Comparison with observations}
\label{obs-theo}
\begin{figure}
\centering
\includegraphics[width=9.0cm, height=11.0cm]{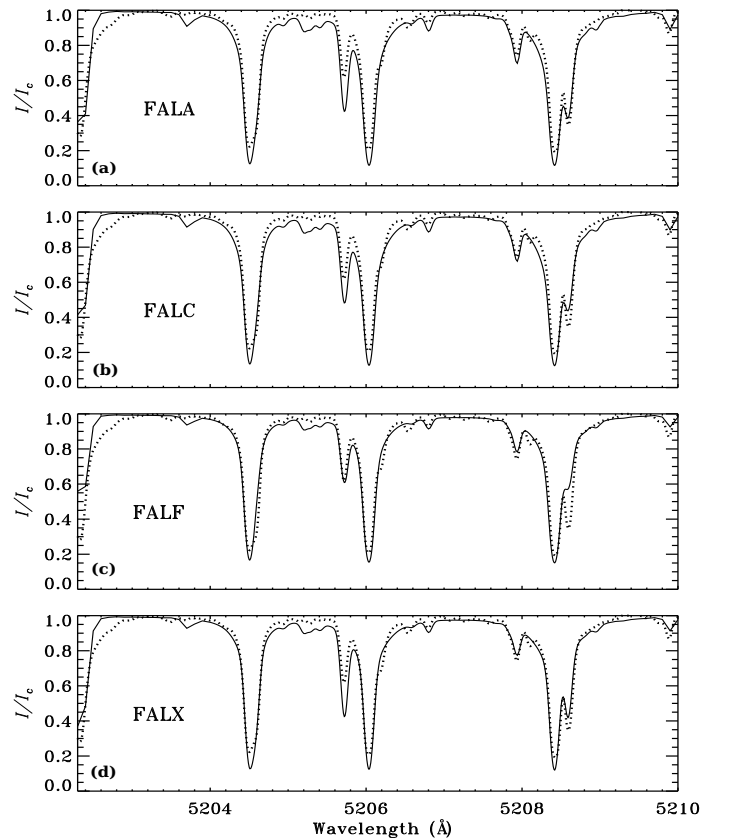}
\caption{ Intensity spectra for a choice of model atmospheres.
The dotted line represents observations and the solid line
represents the theoretical profiles. {The line of sight 
is represented by $\mu=0.15$}.}
\label{int-all}
\end{figure}

\begin{figure}
\centering
\includegraphics[width=9.0cm, height=11.0cm]{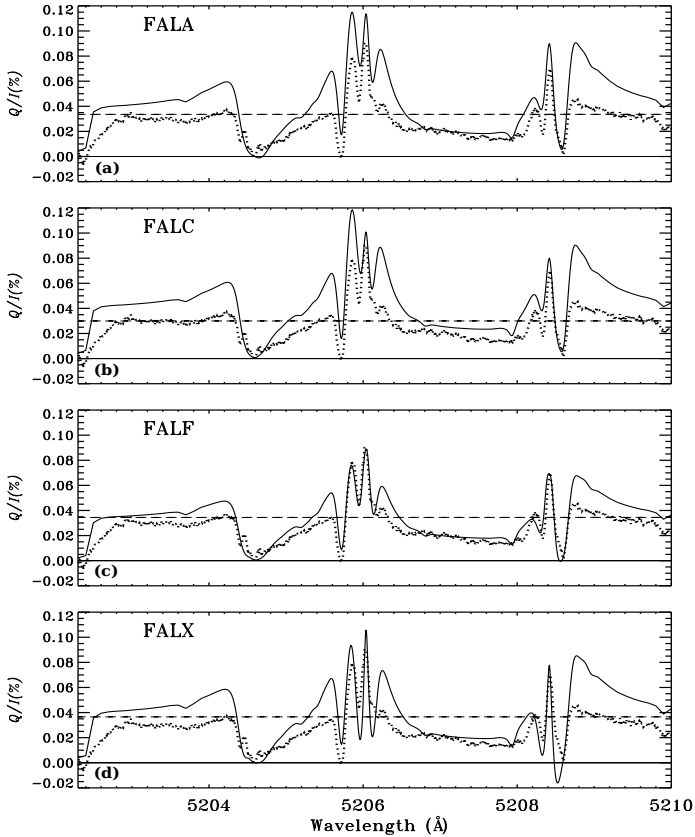}
\caption{ The polarized ($Q/I$) spectra {computed at $\mu=0.15$
and} for a choice of model atmospheres. The line types are the same as in
Fig.~{\ref{int-all}}. The dashed line represents the level 
of continuum polarization. {The observations are taken at $\mu=0.15$.
The $B_{\rm turb}$ values used for the theoretical profiles 
are given in Table~{\ref{table-1}}}.}
\label{pol-all}
\end{figure}

In this section we compare the theoretical Stokes profiles computed using
several standard atmospheric models of the Sun, with the observations.
Figs.~{\ref{int-all}} and {\ref{pol-all}} show the $I/I_c$ and $Q/I$
spectra. From Fig.~{\ref{int-all}} we can see that the $I/I_c$ is not
sensitive to the choice of the model atmospheres, whereas $Q/I$
is very sensitive. The reason for this sensitivity is the angular 
anisotropy of the radiation field, which is different for
different atmospheres.

%-------------------------------------------------------
%------------------------------------------------------

From Fig.~{\ref{clv-temp}} it is clear that the temperature structure
of these models are considerably different from each other in the line
formation region. We find that a modification of the temperature structure at
certain range in the atmosphere does not significantly affect the emergent $I/I_c$
profile. However $Q/I$ is quite sensitive to such `modifications' in the 
temperature structure in the line formation region. The theoretical profiles (solid lines)
in Fig.~{\ref{pol-all}} are computed taking into account
the effects of microturbulent magnetic fields ($B_{\rm turb}$) with an isotropic angular 
distribution (Stenflo 1994). Further, the spectral smearing (see Anusha et al. 2010)
is done using a Gaussian function with FWHM of 80\,m\AA. The use of
 $B_{\rm turb}$ is essential to obtain correct line center amplitudes
of $Q/I$. The values of $B_{\rm turb}$ for the three components of the Cr\,{\sc i}
triplet are different. They are chosen to fit the observed line center amplitudes
of $Q/I$ using the FALF model. {In this way, the microturbulent magnetic 
field strength is only used as a free parameter to improve the fit 
with the observations.}
 We have made no attempt to achieve a good fit
to the line center amplitudes of the $(Q/I)$ profile computed
using FALA, FALC and FALX models. FALF provides a better fit to the
observed $Q/I$ profiles at the cores of the three lines and the interference regions
in between them. However the far
wings still remain poorly fitted even by the FALF model. {To achieve a 
good fit to the far wing region of all the three lines,} we found
it necessary to enhance (see Fig.~{\ref{clv-temp}}) the temperature in the layers
where the far wings are formed.

\begin{figure}
\centering
 \includegraphics[width=9.0cm, height=8.0cm]{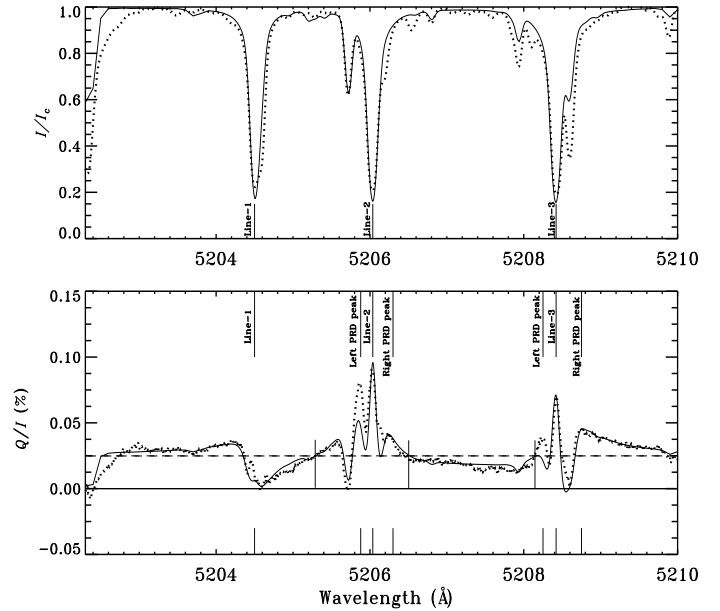}
\caption{\footnotesize Comparison between the {limb ($\mu=0.15$)}
observations (dotted line,
 representing the $Q/I$ spectrum of Fig.~{\ref{observations}} averaged
 along the slit) and the 
theoretical profile (solid line).
The centers of the 3 lines, the PRD peak positions of line-2 and line-3, and
the cross-over wavelength positions between the lines are marked with vertical lines.
The dashed line in $Q/I$ at 
0.025\,\%\ represents the continuum polarization level.
The solid line is the same as the dashed line of Fig.~{\ref{crd-prd}a},
except that we have now introduced Hanle depolarization
due to microturbulent magnetic fields {(see Table~{\ref{table-1}})}, and added spectral
smearing  of 80\,m\AA\ to simulate the observations. The smearing wipes out the
double-peaked core structure that we see in Fig.~{\ref{crd-prd}a}.}
\label{fit}
\end{figure}
%%%%%%%%%%%%%%%%%%%%%%%%%%%%%%%%%%%%%%%%%%%%%%%fig3

Fig.~{\ref{fit}} shows a comparison between the 
profiles of the Cr\,{\sc i} triplet
computed with the $J$-state interference theory (solid line) { and the observed data (dotted line).} 
This solid line is the same as the dashed 
line in Fig.~{\ref{crd-prd}a}, except that it now also includes 
the contributions from $B_{\rm turb}$
and a spectral smearing of 80 m\AA\  to simulate the observations.
The best fit values of 
$B_{\rm turb}$ for the 3 components  of the Cr\,{\sc i}
triplet are given in Table~{\ref{table-1}}.
The approximate heights of formation given in Table~{\ref{table-1}}
are the heights at which the condition $\tau(\lambda_0)/\mu \simeq 1$
is satisfied for $\mu=0.15$. The quantity $\tau(\lambda_0)$
is the total optical depth at line center { for the $\overline{\rm FALF}$ 
model}.

The observed $Q/I$ spectra of the Cr\,{\sc i} triplet have two main characteristics, namely
(i) the presence of a triple peak structure in line-2 and line-3; (ii)
the cross over of $Q/I$ about the continuum polarization level, occurring between
the line components. 
These two aspects are well reproduced in terms of the theoretical
framework with the 
redistribution matrix theory for $J$-state interference
developed in P1 and P2. The small discrepancies
between the observations and theoretical profiles in $Q/I$ can
be attributed to the presence of blend lines. The blend lines are 
assumed to be formed under LTE and generally depolarize the wings 
of the main line as well as the continuum polarization.
To fit the observed $I$ spectra the oscillator strengths of the 
blend lines from the Kurucz data base are used unchanged, with
the single exception of the Y\,{\sc ii} line at 5205.75\,\AA, since 
the value from the data base for this line does not
reproduce the Stokes $I$ spectrum at all. Therefore the oscillator
strength for this line is changed
substantially to get a good
Stokes $I$ fit. As soon as the Stokes $I$ fit becomes good, the $Q/I$
fit automatically becomes good as well { around 5205.75\,\AA. Such
enhancement of the Y\,{\sc ii} line oscillator strength
is also used in computing the theoretical profiles
shown in Figs.~{\ref{int-all}} and {\ref{pol-all}}}. 
The discrepancy in the theoretically computed and observed intensity spectra 
of other blend lines is slightly model atmosphere dependent,
particularly for the one at 5208.6\ \AA\ { (see Fig.~{\ref{int-all}} for details)}.
We have not made a detailed 
attempt to simultaneously fit the $(I,Q/I)$ spectra of all the blend lines.

The reasons for the lack of a good fit to the observed $Q/I$ at the left-wing peaks of
line-2 and line-3 remain unclear and need further investigation.
The deviations of the model fit from the observations are possibly
due to unidentified opacity sources. These deviations however do not
affect the diagnostic potential of the Cr\,{\sc i} triplet.

%===========Table==============================================================
\begin{table}
\caption{{Microturbulent magnetic field strengths necessary 
to obtain a best fit to the line center value of the observed $Q/I$
using $\overline{\rm FALF}$.}} 
\label{table-1}
\centering
\begin{tabular}{|ccc|}
 \hline
Line & {height at which $(\tau_{\lambda}/\mu)=1$} & $B_{\rm turb}$  \\
 & {for $\mu=0.15$} & (G)\\
\hline \hline
5204.50\ \AA  & 845 km &  4.0  \\        
5206.04\ \AA  & 884 km &  6.0  \\
5208.42\ \AA  & 953 km &  4.5 \\
\hline
\end{tabular}
\end{table}

%======================================================================================
\section{Conclusions}
\label{conclusions}
In the present paper we have studied the importance of $J$-state interference 
phenomena with realistic radiative transfer
modeling of the Second Solar Spectrum. We have selected the Cr\,{\sc i}
triplet for this purpose and made use of the theory for partial 
frequency redistribution with  $J$-state interference developed in P1 and P2
in the absence of lower term polarization. This theory is used in combination with
realistic atmospheres and a model atom for Cr\,{\sc i}.
Our results demonstrate that it is indeed possible
to obtain an excellent fit to the observed $Q/I$ profiles 
without use of lower term polarization, but they also clearly show that
accounting for the PRD mechanism is essential to model the 
observed scattering polarization in sufficient detail. The CRD approach alone
cannot be used to model the observed spectra. 
 {We note that Belluzzi and Trujillo Bueno (2011)
have carried out a basic investigation of the $J$-state interference phenomenon on
different multiplets, neglecting RT and PRD effects (the theory they apply is based
on the CRD assumption). Nevertheless, they were able
to identify and explain qualitatively the observational signatures produced by
$J$-state interference in the Cr\,{\sc i} triplet (i.e., the cross over of $Q/I$ about the
continuum level occurring between the lines, and the $Q/I$ feature around the line-1
core), neglecting and including the effects of lower-term polarization
and dichroism.}

Our observations were performed in non-magnetic
regions, but we find 
that microturbulent magnetic fields with an isotropic
angular distribution are needed to fit the line center amplitudes
of the $Q/I$ spectra. 

The near wing PRD peaks and the 
characteristic cross-overs in $Q/I$ that are typical of $J$-state interference 
are well modeled only through a weighted combination of partially coherent 
(through $R_{\rm II}$) and completely non-coherent (through CRD) scattering processes. 
The weighting factors (branching ratios) are the ones used to represent the collisional 
frequency redistribution in line scattering, and they are 
properly accounted for in our RT calculations. We find that elastic collisions
indeed play a major role in modeling the wing polarization of the Cr\,{\sc i}
triplet.
A hotter model atmosphere (FALF) with a slight additional temperature enhancement is
found to be needed 
to obtain a good fit to the observed data, in particular for $Q/I$. This
{emphasizes} that the $Q/I$ spectrum (together with the $I$ spectrum) provides a
much stronger constraint on the model atmosphere than the intensity spectrum alone.
The Second Solar Spectrum is thus not only useful for magnetic field diagnostics,
but also for modeling the thermodynamic structure of the Sun's atmosphere.

%==================================================================================
%%\acknowledgment
\begin{acknowledgements}
\noindent
{We thank the Referee for constructive suggestions, which helped to
considerably improve the paper.}
HNS thanks the Indo-Swiss Joint Research Program
of SER and DST, for supporting her visit to IRSOL. Research at IRSOL is 
financially supported by Canton Ticino, the foundation Aldo e Cele Dacc\`o, 
the city of Locarno, the local municipalities and the SNF grant 200020-127329. 
R.R. acknowledges the foundation Carlo e Albina Cavargna for financial 
support. The authors thank Dr. R. Holzreuter and  Dr. H. Frisch for useful discussions.
The authors like to thank Dr. V. Bommier for providing the
programs to compute type-III redistribution functions and the quadrature for angle integrations.
The authors are grateful to Dr. Dominique Fluri, who kindly
provided a version of the realistic line modeling code which is generalized
in the present paper to include $J$-state interference phenomenon.
{The authors would like to thank Dr. Han Uitenbroek for providing
a version of his RH-code.}
\end{acknowledgements}

%========== References ==============================================

\end{document}